\documentclass[sigconf,natbib=true]{acmart}
\usepackage{booktabs}
\usepackage{multirow}
\usepackage{enumitem} 

\usepackage{amsthm,amsmath}
\usepackage{mathrsfs}
\usepackage{cleveref}
\usepackage{enumitem}
\usepackage{multicol}
\usepackage[table,xcdraw]{xcolor}

\usepackage{graphicx}
\usepackage{booktabs}
\usepackage{array}
\usepackage{tabularx}

\AtBeginDocument{%
  }

\copyrightyear{2026}
\acmYear{2026}
\setcopyright{cc}
\setcctype{by-nc-nd}
\acmConference[KDD '26]{Proceedings of the 32nd ACM SIGKDD Conference
on Knowledge Discovery and Data Mining V.1}{August 09--13, 2026}{Jeju
Island, Republic of Korea}
\acmBooktitle{Proceedings of the 32nd ACM SIGKDD Conference on Knowledge
Discovery and Data Mining V.1 (KDD '26), August 09--13, 2026, Jeju Island,
Republic of Korea}
\acmPrice{}

\begin{document}

\title{Unleashing the Potential of Neighbors: Diffusion-based Latent 
\\ Neighbor Generation for Session-based Recommendation}

\author{Yuhan Yang}
\affiliation{%
  \institution{University of Electronic Science and Technology of China}
  \city{Chengdu}
  %\state{Sichuan}
  \country{China}
}
\email{y.yuhan2000@gmail.com}

\author{Jie Zou}
\authornote{Corresponding author.}
\affiliation{%
  \institution{University of Electronic Science and Technology of China}
  \city{Chengdu}
  %\state{Sichuan}
  \country{China}
}
\email{jie.zou@uestc.edu.cn}

\author{Guojia An}
\affiliation{%
  \institution{University of Electronic Science and Technology of China}
  \city{Chengdu}
  %\state{Sichuan}
  \country{China}
}
\email{an_guojia@163.com}

\author{Jiwei Wei}
\affiliation{%
  \institution{University of Electronic Science and Technology of China}
  \city{Chengdu}
  %\state{Sichuan}
  \country{China}}
\email{mathematic6@gmail.com}

\author{Yang Yang}
\affiliation{%
  \institution{University of Electronic Science and Technology of China}
  \city{Chengdu}
  %\state{Sichuan}
  \country{China}}
\email{yang.yang@uestc.edu.cn}

\author{Heng Tao Shen}
\affiliation{%
  \institution{University of Electronic Science and Technology of China}
  \city{Chengdu}
  %\state{Sichuan}
  \country{China}}
\email{shenhengtao@hotmail.com}

\renewcommand{\shortauthors}{Yuhan Yang et al.}

\begin{abstract}
Session-based recommendation aims to predict the next item that anonymous users may be interested in, based on their current session interactions. Recent studies have demonstrated that retrieving neighbor sessions to augment the current session can effectively alleviate the data sparsity issue and improve recommendation performance. 
However, existing methods typically rely on explicitly observed session data, neglecting latent neighbors - not directly observed but potentially relevant within the interest space - thereby failing to fully exploit the potential of neighbor sessions in recommendation. 

To address the above limitation, we propose a novel model of diffusion-based latent neighbor generation for session-based recommendation, named \textbf{DiffSBR}. 
Specifically, DiffSBR leverages two diffusion modules, including retrieval-augmented diffusion and self-augmented diffusion, to generate high-quality latent neighbors. In the retrieval-augmented diffusion module, we leverage retrieved neighbors as guiding signals to
constrain and reconstruct the distribution of latent neighbors. Meanwhile, we adopt a training strategy that enables the retriever to learn from the feedback provided by the generator. In the self-augmented diffusion module, we explicitly guide the generation of latent neighbors by injecting the current session's multi-modal signals through contrastive learning. 
After obtaining the generated latent neighbors, we utilize them to enhance session representations for improving session-based recommendation. Extensive experiments on four public datasets show that DiffSBR generates effective latent neighbors and improves recommendation performance against state-of-the-art baselines.

\end{abstract}




\maketitle

\section{Introduction}
Given the exponential growth of multimedia content, users face increasing challenges in finding information that aligns with their preferences \cite{mb52,mb54}. 
Fortunately, the emergence of recommender systems \cite{mb43,mb57,mb41,mb51,mb55} has alleviated this issue. However, due to growing concerns over user privacy, it is often infeasible to access users' personal information and historical interactions. To address this issue, session-based recommendation (SBR) \cite{mb45,mb05,mb46} has emerged as a promising solution. SBR aims to recommend the next item based solely on the interaction of anonymous users within a session.

In SBR, the lack of user profiles, coupled with short session lengths, intensifies data sparsity, thereby degrading overall recommendation performance. To alleviate this issue, recent studies primarily adopt retrieval-based neighbor methods \cite{mb06,mb10} to mine neighbor information and enhance the target session representation. These methods can be broadly categorized into similarity-based and co-occurrence-based approaches. The former (e.g., ICM-SR \cite{mb08}, DIDN \cite{mb07}, TASI-GNN \cite{mb11}, ECCL \cite{mb03}) select semantically similar sessions via static similarity learning, while the latter (e.g., FGNN \cite{mb22}, MSGAT \cite{mb06}, DGNN \cite{mb23}, DIMO \cite{mb10}) leverage item co-occurrence patterns to capture structural relationships. 

\begin{figure}[t]
	\begin{center}
		\includegraphics[width=0.95 \linewidth]{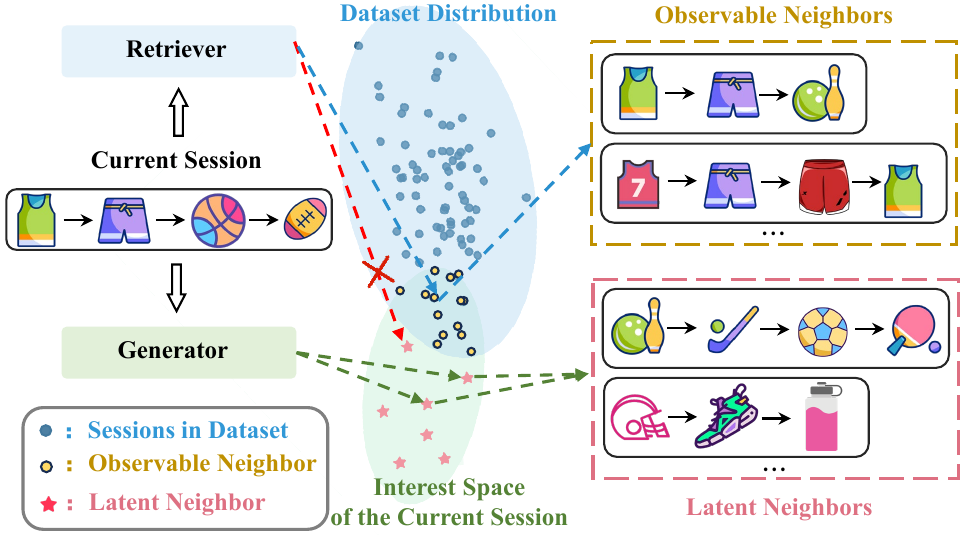}
	\end{center}
	\caption{Retrieval-based method vs. generative method. The blue region represents the distribution of the dataset, while the green region denotes the full interest space of the current session. Retrieval-based method can only access interest-aligned neighbors within the intersection of these regions, thus being constrained by observed data. In contrast, the generative method overcomes this limitation by exploring the entire interest space, enabling the generation of potentially relevant but unobserved latent neighbors.}
	\label{fig: f1}
\end{figure}

Although these existing neighbor retrieval methods have made notable progress in the field of SBR, they still face inherent limitations. These methods are constrained by the scope of the dataset and only retrieve existing \textit{\textbf{observable neighbors}, which refer to interest-aligned sessions explicitly contained in the dataset}. However, this paradigm overlooks \textit{\textbf{latent neighbors}, which are not recorded in the dataset and therefore cannot be discovered through retrieval methods but are still aligned with the user’s underlying interests}, thereby limiting the expressiveness of neighbors and ultimately constraining recommendations. Specifically, as shown in Figure \ref{fig: f1}, we first project the user's current session representation into a unified interest space. Under the retrieval paradigm, observable neighbors are obtained from samples that exist in the dataset (blue region in Figure \ref{fig: f1}). While such retrieved neighbors may offer partial coverage of the user’s interest space, they are fundamentally constrained by the boundaries of the dataset, thereby failing to access the full extent of the user’s interest space. In contrast, the generative paradigm models the potential distribution of user interests and expands the scope of neighbors (within the green region but outside the green–blue overlap in Figure \ref{fig: f1}). This approach enables active exploration of interest regions beyond the observed data, allowing the generation of latent neighbors that are inaccessible through traditional retrieval methods. 

To overcome the inherent limitations of the aforementioned retrieval paradigm, we take an initial step toward transitioning from a retrieval-based framework to a generative paradigm, aiming to generate latent neighbors beyond the scope of observable neighbors. Inspired by the remarkable advances of diffusion models in modeling complex data distributions \cite{mb49,mb38}, we adopt diffusion as the main generative mechanism in our framework. Despite their promising generative capabilities, diffusion models still face two key challenges in ensuring the quality and effectiveness of the generated latent neighbors: (1) How to effectively design guidance mechanisms to improve the quality of latent neighbors generated by diffusion models. (2) How to effectively integrate multi-modal information to enhance latent neighbors generation in the diffusion generation process.

To address the aforementioned two challenges, we propose a novel model of \textbf{Diff}usion-based latent neighbor generation for \textbf{S}ession-\textbf{b}ased \textbf{R}ecommendation, called \textbf{DiffSBR}, which guides the diffusion model to generate high-quality latent neighbors. Specifically, to tackle the first challenge, we proposed a novel module called \textbf{Retrieval-augmented Diffusion Module}, which controls generation by incorporating the retrieved prior neighbor information during the diffusion generation process, so that latent neighbors can be generated in a targeted manner. Moreover, a training strategy is introduced in this module to reversedly optimize the neighbor retriever using the loss signal from the neighbor generator, establishing a closed-loop collaboration between retrieval and generation. To address the second challenge above, we design a \textbf{Self-augmented Diffusion Module}, which integrates multi-modal information into the process of latent neighbor generation. Specifically, we use multi-modal information to guide generation and then conduct contrastive learning against the retrieval-augmented diffusion outputs to better integrate multi-modal semantics.

In this paper, our main contributions are summarized as follows:

\begin{itemize}
\item {We perform a pioneering attempt to generate latent neighbors and highlight their importance, aiming to mitigate the inherent limitations of traditional retrieval-based methods in SBR.}
\item {We propose a novel DiffSBR model, an effective framework that generates effective latent neighbors through a retrieval-augmented diffusion module and a self-augmented diffusion module, thereby improving recommendation performance.}

\item {We conduct extensive experiments on four public datasets, demonstrating that our proposed DiffSBR model, not only significantly outperforms existing SBR methods, but also proves the effectiveness and necessity of generating latent neighbors.}
\end{itemize}

\section{Related Work}
\subsection{Session-based Recommendation}
SBR models user preferences from short-term anonymous interactions. Early methods like FPMC \cite{mb19} combine Markov chains with matrix factorization but fail to capture high-order dependencies. GRU4Rec \cite{mb20} and NARM \cite{mb14} improve sequential modeling via RNNs and attention. Recent GNN-based models (e.g., SR-GNN \cite{mb15}, TAGNN \cite{mb21}) treat sessions as graphs to better capture complex item relations. However, they rely solely on intra-session information, ignoring valuable signals from neighboring sessions. To mitigate this, some methods retrieve neighbor sessions to enrich current session representations.

\textbf{Retrieval-based neighbor methods in SBR.} These methods aim to integrate cross-session neighbor information to enhance session representations under sparse interactions. Existing studies can be broadly categorized into two main types. The first focuses on similarity-based retrieval methods \cite{mb03,mb06}, which identify relevant neighbor sessions by computing the similarity. For example, ICM-SR \cite{mb08} employs an intention-guided neighbor detector to locate relevant sessions, while DIDN \cite{mb07} utilizes a dynamic intention-aware module to retrieve semantically similar sessions. The second line of work leverages co-occurrence relationships \cite{mb10,mb12}, constructing global graphs based on item co-occurrence to capture pairwise transitions across sessions. For instance, CGL \cite{mb09} builds a global graph to model inter-session correlations to enhance item representations. MSGAT \cite{mb06} further constructs a session-level relation graph and incorporates an intent-aware collaboration module to refine the session representation. 

Although the aforementioned retrieval-based methods can achieve effectiveness, they only retrieve observable neighbors from the dataset, which consequently limits the quality of the neighbors. In contrast, we exploit the generative capability of diffusion models to generate latent neighbors, thereby uncovering semantically relevant neighbors that are not explicitly present in the data.

\textbf{Multi-modal-based methods in SBR.} Since user interests are often driven by multi-modal content, relying solely on ID features is insufficient to reflect true preferences. Recent works leverage rich item features to improve user intent modeling. For example, CoHHN \cite{mb13} incorporates price as a key modality; MMSBR \cite{mb04} combines item text and images; LLM4SBR \cite{mb24} utilizes textual descriptions and prompts large language models for intent inference; and DIMO \cite{mb10} decouples and fuses ID and multi-modal features. Distinct from these approaches, our method is the first to explicitly integrate multi-modal signals into the neighbor generation process in SBR, to the best of our knowledge.

\begin{figure*}[t]
	\begin{center}
\includegraphics[width=1\linewidth]{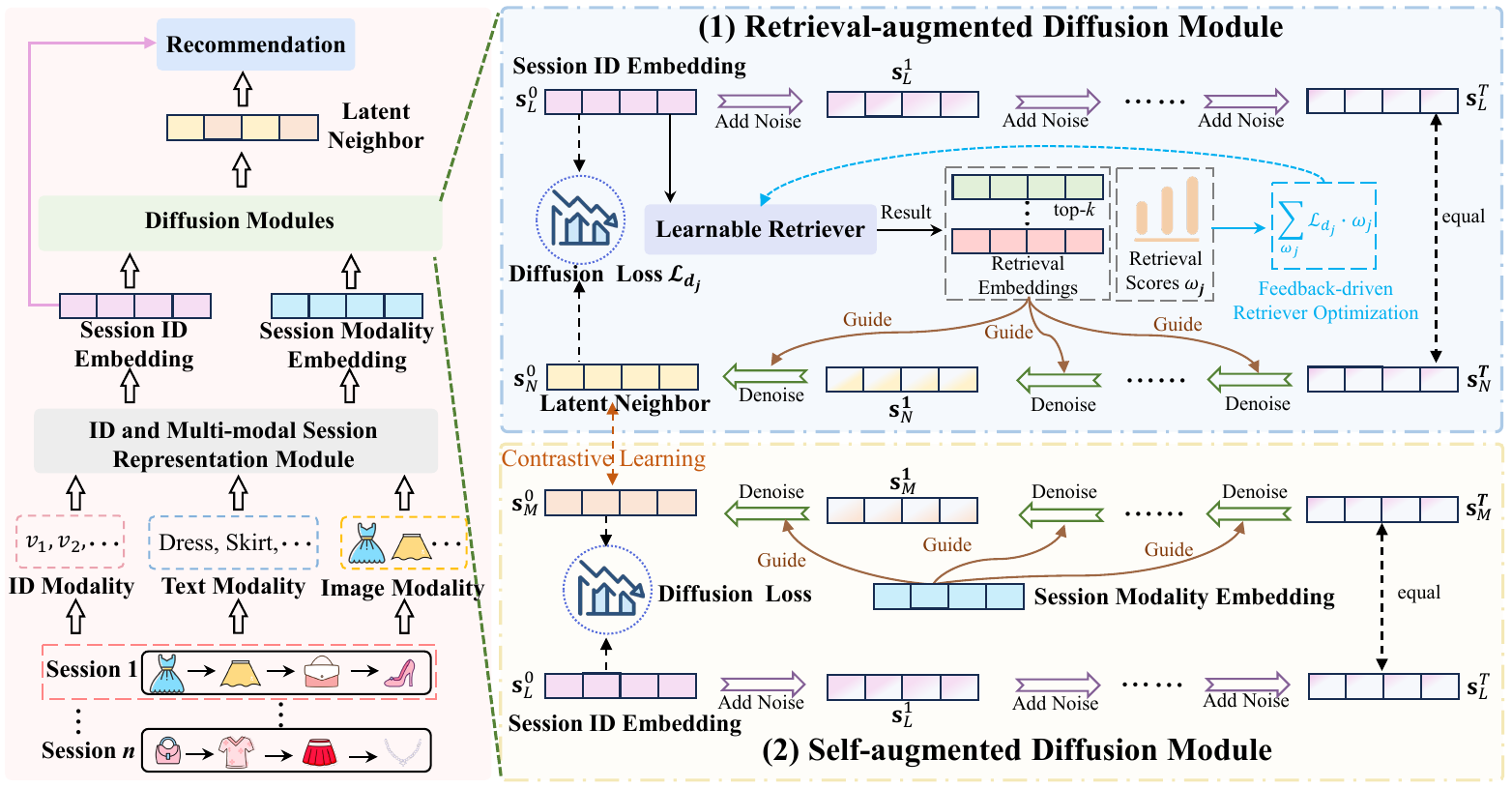}
		
	\end{center}
	\caption{The overall framework of DiffSBR. The left part is the overall framework, and the right part represents the two main components of our diffusion module: (1) the retrieval-augmented diffusion module, which generates latent neighbors guided by retrieved priors; (2) the self-augmented diffusion module, which leverages the session's own multi-modal information to improve the quality of latent neighbors.}

	\label{fig: f2}
\end{figure*}

\subsection{Diffusion Models in Recommendation}
Diffusion models have recently emerged as a promising generative framework for recommendation, offering strong capacity for uncertainty modeling and flexible preference generation. Early work like DiffRec \cite{mb25} applies diffusion to user preference modeling and item generation. DiffuRec \cite{mb26} and DiQDiff \cite{mb02} extend this idea to sequential recommendation via reverse diffusion. Further, DCASR \cite{mb50} and DiffuASR \cite{mb28} explore diffusion-based augmentation for improving sequence representation. Beyond sequences, DiffKG \cite{mb29} and DiffMM \cite{mb01} introduce diffusion into structured data, including knowledge graphs and multimodal interaction graphs. Recently, DDRM \cite{mb30} and MCDRec \cite{mb31}, employ conditional diffusion to integrate user preferences into the generation process, enhancing personalization and semantic alignment. Although prior efforts have validated the effectiveness of diffusion models, most methods focus on target item generation or sequence augmentation. Instead, in this paper, we take the first step to generate latent neighbor information in SBR, which remains largely underexplored by previous work.

\section{Problem Formulation}

Let $\mathcal{V}=\left\{v_{1}, \ldots, v_{i}, \ldots, v_{n}\right\}$ denote the set of all items in the dataset, where $n$ is the total number of unique items. 
Each item $v_i \in \mathcal{V}$ consists of an identifier $v_i^{id}$ and multi-modal content features $v_i^{mo}$, i.e., $v_{i}=\left\{v_{i}^{i d}, v_{i}^{m o}\right\}$. In this work, the multi-modal content features $v_i^{mo}$ include textual and visual modalities: $v_{i}^{m o}=\left\{v_{i}^{t x t}, v_{i}^{i m g}\right\}$, where $v_i^{txt} = \left\{w_1, w_2, \dots, w_q\right\}$ denotes a sequence of $q$ words describing the item's title and brand name, and $v_i^{img}$ represents the corresponding image of item $v_i$. Let $\mathcal{S} = \left\{s_1, \dots, s_j, \dots, s_{|\mathcal{S}|}\right\}$ denote the set of all sessions, where $|\mathcal{S}|$ is the total number of sessions. Each session $s_j$ is an ordered sequence from an anonymous user within a short time period, formally defined as:
$S_{j}=\left[v_{1}, v_{2}, \ldots, v_{m}\right]$, where $m$ is the length of the current session. Given the current session history $\left[v_{1}, v_{2}, \ldots, v_{m}\right]$, the objective of SBR is to predict the next item $v_{m+1}$ that the user is most likely to interact with.

\section{Methodology}
In this section, we provide a detailed description of our proposed DiffSBR framework, as illustrated in Figure~\ref{fig: f2}. 
Each input session is first encoded via an ID and a multi-modal session representation module, resulting in a session ID embedding and a session modality embedding. These representations are then fed into the retrieval-augmented diffusion module and the self-augmented diffusion module to generate latent neighbors, thereby enhancing the current session for the final recommendation task.

\subsection{ID and Multi-modal Session Representation}
Here, we aim to obtain the session ID embedding and session modality embedding for each session separately through three main steps. 

\subsubsection{Initialization of Item Embeddings}
Given the differences in the presentation of various modalities, we adopt specific encoding methods to convert raw modality data into vector representations. For each item $v_i$, we construct three types of embeddings: a structured ID embedding, a semantic embedding from textual descriptions, and a visual embedding derived from item images.

For structured ID embedding $v_i^{id}$ of each item $v_i$, 
we follow common practices in prior work (e.g., \cite{mb14,mb15}) to construct an ID embedding table $\mathbf{E}^{id} \in \mathbb{R}^{n \times d}$ , where each row corresponds to a randomly initialized ID embedding  $\mathbf{e}_i^{id} \in \mathbb{R}^d$ of a specific item.

There exists a substantial semantic gap between textual and visual modalities. Similar to \citet{mb10}, we address this issue by transforming the visual modality into a textual form, thereby aligning heterogeneous modalities into a shared semantic space. Specifically, we employ GoogLeNet \cite{mb35} to predict the top-2 category labels for each image of items, which are concatenated to form a pseudo-textual representation $v_i^{ctxt} = \{ w_1', w_2', \ldots, w_o' \}$. This is further concatenated with the original item description $v_i^{txt} = \{ w_1, w_2, \ldots, w_q \}$ to construct a unified multi-modal input $v_i^{mo} = \{ w_1, \ldots, w_q, w_1', \ldots, w_o' \}$. We feed $v_i^{mo}$ into a pre-trained BERT \cite{mb36} model to obtain contextualized token embeddings $\{ \mathbf{e}_1, \mathbf{e}_2, \ldots, \mathbf{e}_{l+o} \}$, and apply average pooling to initialize the multi-modal item representation $\mathbf{e}_{i}^{m o} \in \mathbb{R}^{d}$, computed as:
\begin{equation}
    \mathbf{e}_{i}^{m o}=\frac{1}{q+o} \sum_{r=1}^{q+o} \mathbf{e}_{r} .
\end{equation}

\subsubsection{Graph-enhanced Item Embeddings}
After obtaining the initialized item embeddings, we incorporate graph structure to further model transition dependencies and collaborative relationships. Following prior work \cite{mb10,mb37}, we construct a directed item graph $\mathcal{G}=(\mathcal{V}, \mathcal{E})$ based on co-occurrence patterns, where each node represents an item and the edge weight between two items reflects their co-occurrence frequency. From this graph, we obtain the corresponding adjacency matrix $\mathbf{A} \in \mathbb{R}^{n \times n}$, where $\mathbf{A}(i, j)$ denotes the edge weight from item $v_i$ to item $v_j$. We also construct the degree matrix $\mathbf{D} \in \mathbb{R}^{n \times n}$, where $\mathbf{D}(i, i)$ is the out-degree of item $v_i$, defined as the sum of its outgoing edge weights.

We then apply an $L$-layer Graph Convolutional Network (GCN) to perform message aggregation. For both ID and multi-modal embeddings (denoted as $\mathbf{e}_i^{id}$ and $\mathbf{e}_i^{mo}$), the $l$-th layer update is defined as:
\begin{equation}   \mathbf{x}_{v_{i}^{c}}^{(l)}=\operatorname{Norm}\left(\mathbf{D}^{-1} \mathbf{A}  \mathbf{x}_{v_{i}^{c}}^{(l-1)} \mathbf{W}^{(l-1)}\right), 
\end{equation}
where $c \in\{i d, m o\}$, $ \mathbf{x}_{v_{i}^{c}}^{(l)}$ denotes the item embedding at the $l$-th GCN layer, and $\mathbf{W}^{(l-1)} \in \mathbb{R}^{d \times d}$ is a learnable transformation matrix. $\text{Norm}(\cdot)$ represents a normalization function. The input to the GCN is the initialized item embedding: $ \mathbf{x}_{v^{id}_{i}}^{0}=\mathbf{e}_{i}^{i d}$, $\mathbf{x}_{v^{mo}_{i}}^{0}=\mathbf{e}_{i}^{m o}$. Finally, we aggregate the representations across all GCN layers to obtain the final graph-enhanced item embedding $ \mathbf{x}_{v_{i}^{c}}$:
\begin{equation}
     \mathbf{x}_{v_{i}^{c}}=\frac{1}{L+1} \sum_{l=0}^{L}  \mathbf{x}_{v_{i}^{c}}^{(l)}.
\end{equation}
\subsubsection{Session Representation}
To further obtain session-level representation, we first employ a contrastive alignment to encourage consistency between id and mo embeddings, which stabilizes multi-modal training, then we compute relevance scores over all items in the sequence, referring to \citet{mb15}. To account for the varying importance of item embeddings, we introduce an attention-based aggregation \cite{mb56} that adaptively weighs each item. The attention weights $\alpha_i$ are computed as:
\begin{equation} \alpha_{i} = \sigma\left(\mathbf{W}_1 \mathbf{x}_{v_{m}^{c}} + \mathbf{W}_2 \mathbf{x}_{v_{i}^{c}}\right), \end{equation}
where $\mathbf{W}_1$ and $\mathbf{W}_2$ are learnable parameters, $\mathbf{x}_{v_{m}^{c}}$ is the last-clicked item in the sequence, and $\sigma(\cdot)$ denotes the sigmoid activation function. Then, the final session representation $\mathbf{s}_{c}$ is obtained by aggregating item embeddings with the learned attention weights:
\begin{equation}   \mathbf{s}_{c}=\sum_{i=1}^{m} \alpha_{i} \cdot \mathbf{x}_{v_{i}^{c}}, \end{equation}
where $\mathbf{s}_{c} \in \left \{ \mathbf{s}_{id}, \mathbf{s}_{mo} \right \}$.

\subsection{Retrieval-augmented Diffusion Module}
In this work, we adopt diffusion models as our generative backbone due to their superior training stability and high-quality sample generation compared to Generative Adversarial Networks (GANs) \cite{mb47} and Variational Autoencoders (VAEs) \cite{mb48}, as demonstrated in prior studies \cite{mb38, mb26, mb02}. Moreover, diffusion models have shown promising performance in recommendation tasks, making them a suitable choice for our generative framework. 

To ensure that the diffusion model explores the interest space in a goal-directed manner, and to prevent it from drifting into semantically irrelevant or user-unrelated regions, we design a retrieval-augmented diffusion module. This module leverages retrieved prior knowledge to guide the diffusion process, enabling the generation of results that are not only semantically coherent but also highly aligned with the user's current preferences. 

\subsubsection{Retrieval-based Prior Construction}
To construct a neighbor set that provides prior guidance for the diffusion process, we first introduce a learnable retriever to select the top-$k$ most relevant neighbors to the current session from the historical session database. Specifically, given a session $\mathbf{s}_{id}$, we compute its similarity score using the following equation: 
\begin{equation}
\operatorname{sim}^{D}=\mathcal{F}_{\text {score }}\left(\left[\mathbf{s}_{id} \| \mathbf{s}^{D}\right]\right),
\end{equation}
where $\mathcal{F}_{\text{score}}$ denotes a multi-layer perceptron, and $[\cdot  ||  \cdot]$ represents vector concatenation. Then we select the top‑$k$ candidate sessions $\mathcal{N}_{k}$ with the highest similarity scores. Next, we apply the softmax function to normalize the similarity scores, in order to obtain the attention weight for each neighbor:
\begin{equation}
    \omega_{j}=\frac{\exp \left(\operatorname{sim}^{D}_{j}\right)}{\sum_{r=1}^k \exp \left(\operatorname{sim}_{r}^{D}\right)}, 
\end{equation}
where $j \in \left \{ 1,...,k \right \}$. The neighbor session ${\mathcal{N}_{k}}$, together with their corresponding score weights, is used as conditional information to guide the diffusion model in the downstream process.

\subsubsection{Diffusion-based Neighbor Generation}\label{4.2.2}
After obtaining prior information to guide the diffusion process, we adopt a conditional DDPM \cite{mb33} to train a denoising network, which progressively removes noise during inference to generate conditionally guided latent neighbors. Specifically, during training, a forward Markov chain is constructed to gradually add Gaussian noise to the original data sample, denoted as the clean session embedding \(\mathbf{s}_{id}=\mathbf{s}_L^0\), over \(T\) time steps \(t \in \{1, \ldots, T\}\). This process eventually produces a noisy vector \(\mathbf{s}_L^T\) that approximates a standard normal distribution. Each forward transition is defined as: 
\begin{equation}
        q\left(\mathbf{s}_L^t \mid \mathbf{s}_L^{t-1}\right)=\mathcal{N}\Bigl(\mathbf{s}_L^t; \sqrt{1-\beta_{t}}\,\mathbf{s}_L^{t-1},\, \beta_{t} \mathbf{I}\Bigr),
\end{equation}
 where $\beta_t \in (0, 1)$ is the noise schedule at time step $t$, and $\mathbf{I}$ is the identity matrix. Let $\alpha_t = 1 - \beta_t$ and define the cumulative product as: $\bar{\alpha}_{t}=\prod_{i=1}^{t} \alpha_{t}$. Then, the noisy sample at step $t$ can be directly derived from the clean embedding as:

\begin{equation}
\mathbf{s}_{L}^{t}=\sqrt{\bar{\alpha}_{t}} \mathbf{s}_{L}^{0}+\sqrt{1-\bar{\alpha}_{t}} \boldsymbol{\epsilon}, 
\end{equation}
where $\boldsymbol{\epsilon} \sim \mathcal{N}(0, \mathrm{I})$ \cite{mb32,mb34}. 
Following previous studies \cite{mb02,mb30,mb38}, we do not explicitly predict the added noise during training. Instead, we directly generate the latent neighbor representation ${\mathbf{s}}_N^{0}$ at each timestep under the guidance of the semantic prior $\mathcal{N}_{k}$, enabling the model to generate potential neighbors in a targeted manner:
\begin{equation}
{\mathbf{s}}_{N}^{0}=f_{\theta}\left(\mathbf{s}_{L}^{t}, \mathcal{N}_{k}, t\right),
\end{equation}
where $f_{\theta}(\cdot)$ adopts an MLP architecture as in \citet{mb02} and \citet{mb27}. In our model, the reverse generation process is conceptualized as a conditional Gaussian distribution at each step:
\begin{equation}
p_{\theta}\left({\mathbf{s}}_{N}^{t-1} \mid {\mathbf{s}}_{N}^{t}, \mathcal{N}_{k}\right)=\mathcal{N}\left({\mathbf{s}}_{N}^{t-1} ; \mu_{\theta}\left({\mathbf{s}}_{N}^{t}, \mathcal{N}_{k}, t\right), \Sigma_{\theta}\left({\mathbf{s}}_{N}^{t}, \mathcal{N}_{k}, t\right)\right),
\end{equation}
where $\mu_{\theta}(\cdot)$ and $\Sigma_{\theta}(\cdot)$ are the learnable mean and covariance predicted by the network. Accordingly, the diffusion loss $\mathcal{L}_{d}$ is defined as:
\begin{equation}
\mathcal{L}_{d}=\mathbb{E}_{t, \mathbf{s}_{L}^{0}, \boldsymbol{\epsilon}}\left[\left\|\mathbf{s}_{id}-f_{\theta}\left(\sqrt{\bar{\alpha}_{t}} \mathbf{s}_{L}^{0}+\sqrt{1-\bar{\alpha}_{t}} \boldsymbol{\epsilon}, \mathcal{N}_{k}, t\right)\right\|^{2}\right]
\end{equation}

During inference, we employ a deterministic strategy to generate latent neighbors ${\mathbf{s}}_{N}^{0}$. Specifically, we first apply $T^{'}$ steps of forward corruption to the input session embedding $\mathbf{s}_{id}$ to obtain a noisy initialization. Then, starting from this point, we perform $T^{'}$ steps of reverse denoising conditioned on the retrieved neighbor prior $\mathcal{N}_{k}$. In the deterministic reverse process, variance is omitted and the predicted mean is used directly:
\begin{equation}
{\mathbf{s}}_{N}^{t-1}=\frac{\sqrt{\bar{\alpha}_{t-1}} \beta_{t}}{1-\bar{\alpha}_{t}} f_{\theta}\left({\mathbf{s}}_{N}^{t}, \mathcal{N}_{k}, t\right)+\frac{\sqrt{\alpha_{t}}\left(1-\bar{\alpha}_{t-1}\right)}{1-\bar{\alpha}_{t}} {\mathbf{s}}_{N}^{t},
\end{equation}
where the model $f_{\theta}(\cdot)$ is encouraged to learn session-specific denoising strategies for each input. The final output \({\mathbf{s}}_N^{0}\) is then used as the generated latent neighbor representation.

\subsubsection{Feedback-driven Retriever Optimization}
Our method is designed to enable the retriever to continually learn from the feedback of the generator, so as to retrieve neighbor sessions that are more beneficial for the neighbor generation task. To achieve this, we evaluate the effectiveness of each retrieved neighbor in guiding the diffusion process.  Specifically, given the target session's ID-based representation $\mathbf{s}_{id}$ and a set of retrieved neighbors $\mathcal{N}_{k}$, we compute the generator's loss $\mathcal{L}_d$ through the diffusion process defined in section \ref{4.2.2}. A smaller value of $\mathcal{L}_d$ indicates that the corresponding neighbor contributes more effectively to the generation of latent neighbors. Inspired by \citet{mb39}, we introduce a relative ranking-based supervision mechanism: for two retrieved neighbors $\mathbf{s}^D_i$, $\mathbf{s}^D_j$ $\in \mathcal{N}_{k}$, if the diffusion loss under $\mathbf{s}^D_i$ as the condition is lower than that under $\mathbf{s}^D_j$ (i.e., $\mathcal{L}_{d_i} < \mathcal{L}_{d_j}$), then $\mathbf{s}^D_i$ is considered more useful for generation and should be assigned a higher retrieval score. To optimize the retriever according to the generator's preferences, we propose a training strategy that incorporates this supervision signal. Considering the large size of the candidate pool in practice, we improve efficiency by applying supervision only to the top-$k$ retrieved neighbors. The corresponding loss function $\mathcal{L}_{r}$ is defined as:
\begin{equation}
\mathcal{L}_{r}=\sum_{\omega_j \in \text { top-}k} \mathcal{L}_{d_{j}} \cdot \omega _{j},  
\end{equation}
where $\omega_j$ denotes the softmax-normalized relevance score of neighbor $\mathbf{s}^D_j$ computed by the retriever, and $\mathcal{L}_{d_j}$ is the diffusion loss incurred when using $\mathbf{s}^D_j$ as the generation condition.

\subsection{Self-augmented Diffusion Module}
To effectively incorporate modality-specific information into the diffusion process without interfering with the retrieval-augmented diffusion, we design a self-augmented diffusion module. This module conducts contrastive learning between the retrieval-augmented and self-augmented diffusion paths, thereby enhancing semantic alignment across modalities.

Specifically, we apply the forward diffusion process to the ID representation $\mathbf{s}_{id}$ by adding Gaussian noise, yielding noisy representations $\mathbf{s}_{L}^{t}$ at timestep $t$. During the reverse diffusion process, unlike the retrieval-augmented diffusion module, which performs denoising under the guidance of retrieved neighbors, the self-augmented diffusion module leverages the multi-modal representations of the current input itself to guide the denoising. The denoising is carried out by the network $f_{\psi}(\cdot)$, yielding the denoised embedding ${\mathbf{s}}_{M}^{t}=f_{\psi}\left(\mathbf{s}_{L}^{t}, \mathbf{s}_{mo},t\right)$. Similar to the retrieval-augmented diffusion module, we compute the diffusion loss $\mathcal{L}_{s}$ for the self-augmented module as follows:
\begin{equation}
    \mathcal{L}_{s}=\mathbb{E}_{t, \mathbf{s}_{L}^{0}, \boldsymbol{\epsilon}}\left[\left\|\mathbf{s}_{id}-f_{\psi}\left(\sqrt{\bar{\alpha}_{t}} \mathbf{s}_{L}^{0}+\sqrt{1-\bar{\alpha}_{t}} \boldsymbol{\epsilon}, \mathbf{s}_{mo}, t\right)\right\|^{2}\right].
\end{equation}

Subsequently, to indirectly inject modality-guided signals into the diffusion process, we adopt a contrastive learning strategy to enhance semantic alignment across modalities. Given the denoised embeddings ${\mathbf{s}}_{N}^{t}$ and ${\mathbf{s}}_{M}^{t}$ at timestep $t$, the contrastive loss is defined as $\mathcal{L}_{m}$:
\begin{equation}
\mathcal{L}_{{m}}=-\frac{1}{B} \sum_{i=1}^{B} \log \frac{\exp \left(\text{sim}\left({\mathbf{s}}_{N,i}^{t},{\mathbf{s}}_{M, i}^{t}\right) /{\tau}\right)}{\sum_{j=1}^{B} \exp \left(\text{sim}\left({\mathbf{s}}_{N,i}^{t},{\mathbf{s}}_{M, j}^{t}\right){\tau}\right)},
\end{equation}
where $B$ is the batch size and $\tau$ is a temperature coefficient, 
and $\text{sim}(\cdot,\cdot)$ calculates cosine similarity. 

\subsection{Prediction and Model Optimization}
The final session representation is defined as a weighted combination of the original ID-based representation and the generated neighbor representation:
\begin{equation}
    \mathbf{s}_{f}=\rho \mathbf{s}_{id}+(1-\rho) {\mathbf{s}}_{N}^{0},
\end{equation}
where $\rho \in[0,1]$ is a learnable parameter that balances the contribution of the two components. For each candidate item $\mathbf{x}_{v_{i}^{id}}$, the predicted click probability $\hat{y}_{i}$ is computed as:
\begin{equation}
    \hat{y}_{i}=\mathbf{s}_{f}^{\top} \cdot \mathbf{x}_{v_{i}^{id}}.
\end{equation}
 
The training objective for recommendation is defined as a standard cross-entropy loss $\mathcal{L}_{e}$, formulated as:
\begin{equation}
    \mathcal{L}_{e}(y, \hat{y})=-\sum_{i=1}^{m}\left[y_{i} \log \left(\hat{y}_{i}\right)+\left(1-y_{i}\right) \log \left(1-\hat{y}_{i}\right)\right],
\end{equation}
where $y$ denotes the one-hot encoding vector of the ground-truth item, and $\hat{y}_{i}$ is the predicted probability for the $i$-th item.

The overall training objective $\mathcal{L}$ is defined as a weighted combination of the  $\mathcal{L}_{e}$, $\mathcal{L}_{r}$, $\mathcal{L}_{s}$  and $\mathcal{L}_{m}$: 
\begin{equation}
    \mathcal{L}=\mathcal{L}_{e}+\gamma (\mathcal{L}_{r} + \mathcal{L}_{s})+\delta\mathcal{L}_{m},
\end{equation}
where $\gamma$ and $\delta$ are hyperparameters that control the weights of the diffusion loss and the contrastive loss, respectively.

\section{Experiment}
To demonstrate the effectiveness of DiffSBR, we conduct extensive experiments guided by the following research questions:
\begin{itemize}[leftmargin=1em]
\item \textbf{RQ1}: How does DiffSBR perform compared to existing methods for SBR?
\item \textbf{RQ2}: Does each proposed component contribute positively to the performance of DiffSBR?
\item \textbf{RQ3}: Are the generated latent neighbors more valid than the retrieved observable neighbors from known data?
\item \textbf{RQ4}: How do different hyperparameter settings influence the performance of DiffSBR?
\end{itemize}

\subsection{Experimental Setup}
\subsubsection{Datasets and Evaluation Metrics}
We evaluate our model on four widely used public datasets: \textbf{Cellphones}, \textbf{Sports}, \textbf{Grocery}, and \textbf{Instacart}, following \cite{mb10}. The first three datasets are from different Amazon\footnote{http://jmcauley.ucsd.edu/data/amazon/} categories and have been widely adopted in SBR. Similar to the preprocessing protocols in \citet{mb10, mb13}, sessions are constructed by grouping all user interactions that occur within a single day. The Instacart dataset is a competition dataset released on Kaggle\footnote{https://www.kaggle.com/c/instacart-market-basket-analysis}. To simulate SBR scenarios, following \citet{mb10}, 20\% of transactions with the shortest length are selected. Regarding modality, Amazon datasets provide both textual and visual information for each item, while for Instacart, only textual data is used. Specifically, the text modality includes item titles and brand names. Following \citet{mb14, mb15}, sessions of length 1 and items appearing fewer than five times are removed. The statistical details of all datasets are presented in Table~\ref{tab:datas}. For evaluation metrics, we follow prior studies \cite{mb10,mb14,mb16} to adopt two commonly used evaluation metrics: P@K (Precision at K) and MRR@K (Mean Reciprocal Rank at K), where $\text{K} \in \{10, 20\}$, to evaluate the performance.

\begin{table}[t]
\centering
\caption{Statistics of datasets.}
\label{tab:datas}
\begin{tabular}{lrrrr}
\toprule
\textbf{Datasets} & \textbf{Cellphones} & \textbf{Sports} & \textbf{Grocery} & \textbf{Instacart} \\
\midrule
\#item         & 9,091   & 14,650  & 7,286 & 10,009  \\
\#interaction  & 123,186 & 282,102 & 151,251 & 380,230 \\
\#session      & 40,344  & 90,492  & 43,648  & 88,022\\
avg. length    & 3.05    & 3.12    & 3.47    & 4.32\\
\bottomrule
\end{tabular}
\end{table}

\subsubsection{Baselines}
To demonstrate the effectiveness of DiffSBR, we compare it against a broad range of representative baselines, including traditional and state-of-the-art SBR models, and diffusion-based sequential recommendation approaches: (1) \textbf{SKNN} \cite{mb40} predicts the next item based on retrieving session neighbors with high similarity from historical sessions. (2) \textbf{NARM} \cite{mb14} uses a GRU with attention to model user intent. (3) \textbf{SR-GNN} \cite{mb15} captures complex item transition relationships using GNN. (4) \textbf{MSGIFSR} \cite{mb17} employs GNN to capture user preferences from continuous segments through co-occurrence patterns. (5) \textbf{Atten-Mixer} \cite{mb18} leverages multi-level user intent to perform multi-stage reasoning on item co-occurrence transitions. (6) \textbf{MSGAT} \cite{mb06} enhances the current session by retrieving neighbor based on cosine similarity and co-occurrence relationships. (7) \textbf{MGS} \cite{mb16} leverages item attributes to retrieve similar neighbors and further estimate user preferences. (8) \textbf{MMSBR} \cite{mb04} is the first method in SBR to combine text and images for modeling user intent. (9) \textbf{DIMO} \cite{mb10} uncovers the relationships between co-occurring items and modalities to disentangle the effects of ID and modality. (10) \textbf{DiffuRec} \cite{mb26} adopts diffusion models to handle sequential recommendation, replacing conventional static item embeddings with probabilistic representations. (11) \textbf{DiQDiff} \cite{mb02} uses quantized user sequences as conditions to guide diffusion generation in sequential recommendation.

\renewcommand{\arraystretch}{1.2} 
\setlength{\tabcolsep}{4pt}
\begin{table*}[t]
\centering
\caption{Comparison of different models across datasets and metrics. The best baseline results are underlined. * indicates statistically significant improvement over all baselines ($p$-value $<$ 0.05).}
\label{tab:all}
\resizebox{\textwidth}{!}{
\begin{tabular}{ccccccccccccccccc}
\toprule
\multirow{2}{*}{\textbf{Model}} & \multicolumn{4}{c}{\textbf{Cellphones}} & \multicolumn{4}{c}{\textbf{Sports}} & \multicolumn{4}{c}{\textbf{Grocery}} & \multicolumn{4}{c}{\textbf{Instacart}} \\
\cmidrule(lr){2-5} \cmidrule(lr){6-9} \cmidrule(lr){10-13} \cmidrule(lr){14-17}
& P@10 & MRR@10 & P@20 & MRR@20 & P@10 & MRR@10 & P@20 & MRR@20 & P@10 & MRR@10 & P@20 & MRR@20 & P@10 & MRR@10 & P@20 & MRR@20 \\
\midrule
\textbf{SKNN}         & 14.31 & 8.84  & 16.48 & 9.06  & 31.79 & 24.23 & 33.98 & 24.39 & 40.40 & 27.64 & 42.40 & 27.78 & 6.78 & 2.06 & 11.79 & 2.41\\
\textbf{NARM}         & 15.42 & 12.43 & 16.80 & 12.53 & 35.55 & 33.40 & 36.67 & 33.57 & 45.67 & 40.39 & 47.14 & 40.59 & 8.27 & 3.02 & 12.19 & 3.25 \\
\textbf{SR-GNN}       & 16.36 & 12.96 & 18.11 & 13.09 & 36.31 & 33.36 & 37.69 & 33.66 & 44.33 & 39.44 & 46.24 & 39.64 & 8.96 & 3.27 & 13.00 & 3.64 \\
\textbf{MSGIFSR}      & 17.80 & 12.40 & 21.16 & 12.64 & 36.27 & 30.36 & 39.65 & 30.59 & 45.45 & 38.16 & 48.15 & 38.35 & 11.56 & 3.74 & 16.44 & 4.02 \\
\textbf{Atten-Mixer}  & 19.51 & 14.54 & 22.28 & 14.71 & 37.30 & 33.63 & 39.19 & 33.86 & 47.65 & 40.71 & 49.56 & 40.84  & 8.11  & 3.12  & 11.53  & 3.36 \\
\textbf{MSGAT}        & 17.22 & 13.41 & 20.01 & 13.67 & 37.19 & 33.69 & 38.53 & 33.91 & 45.20 & 39.98 & 47.01 & 40.12 & 9.29 & 3.54 & 13.36 & 3.77 \\
\textbf{MGS}          & 21.54 & 14.24 & 25.02 & 14.48 & 36.79 & 32.39 & 38.45 & 32.50 & 46.59 & 38.83 & 48.37 & 38.98 & 8.95 & 2.87 & 13.74 & 3.09 \\
\textbf{MMSBR}        & 20.59 & 13.94 & 22.82 & 14.13 & 36.69 & 32.52 & 38.29 & 32.73 & 46.05 & 39.01 & 47.89 & 39.23 & 9.89 & 3.61 & 14.37 & 3.84 \\
\textbf{DIMO}         & \underline{31.66} & \underline{16.98} & \underline{38.81} & \underline{17.36} & \underline{45.07} & \underline{34.86} & \underline{49.86} & \underline{35.15} & \underline{53.03} & \underline{41.81} & \underline{57.01} & \underline{41.98}  & \underline{12.51} & \underline{4.31} & \underline{18.36} & \underline{4.81}\\
\textbf{DiffuRec}    & 25.78 & 15.54 & 30.68 & 15.88 & 41.25 & 33.86 & 47.31 & 34.15 & 50.78 & 38.22 & 54.39 & 38.47 & 9.36 & 3.58 & 13.81 & 3.88 \\
\textbf{DiQDiff}     & 28.12 & 16.41 & 33.19 & 16.83 & 43.01 & 34.51 & 49.28 & 34.96 & 52.29 & 39.04 & 56.03 & 39.32 & 10.62 & 3.96 & 15.34 & 4.17 \\
\midrule

\textbf{DiffSBR}      & \textbf{37.28*} & \textbf{18.03*} & \textbf{45.97*} & \textbf{18.60*} & \textbf{49.94*} & \textbf{35.66*} & \textbf{55.83*} & \textbf{36.07*} & \textbf{57.05*} & \textbf{42.64*} & \textbf{62.33*} & \textbf{42.84*} & 
\textbf{13.83*} &
\textbf{5.12*} &
\textbf{20.34*} &
\textbf{5.57*}\\
\textbf{Improvement} $\uparrow$       & 17.75\% & 6.18\% & 18.45\% & 7.14\% & 10.81\% & 2.29\% & 11.97\% & 2.62\% & 7.58\% & 1.99\% & 9.33\% & 2.05\% & 10.55\% & 18.79\% & 10.78\% & 15.80\% \\
\bottomrule         
\end{tabular}}
\end{table*}

\subsubsection{Implementation Details}
Following prior studies \cite{mb04,mb10}, we adopt the Adam optimizer with an initial learning rate of 0.001 and set the mini-batch size to 50. To ensure fair comparison, the embedding dimension for all methods is set to 100. Following \citet{mb10}, we apply PCA to reduce both modalities to 100 dimensions. We perform grid search to select the optimal hyperparameters of the model. The number of GCN layers is set to 3, and the temperature coefficient is set to 0.3. For DDPM, we follow \citet{mb02} by using 32 diffusion timesteps and adopting a truncated linear noise schedule. 

\subsection{Overall Performance (RQ1)}
Table~\ref{tab:all} reports the evaluation results of performance comparison. The results lead to several key observations:

(1) Early methods such as SKNN rely solely on session similarity for neighbor retrieval, but lack the capacity to capture complex item transitions within sessions, resulting in limited recommendation accuracy. Later models like NARM and SR-GNN introduce attention and GNN mechanisms to enhance session modeling, thus improving performance. Recent methods such as MSGIFSR, Atten-Mixer, MSGAT, and MGS attempt to combine neighbor retrieval with graph-based modeling. By integrating historically similar sessions as auxiliary information, these models aim to enrich the session representation, and thereby improve recommendation performance. Nevertheless, they may rely only on observed neighbors and struggle to capture unobserved but semantically relevant sessions. In contrast, DiffSBR introduces a retrieval-augmented diffusion module that treats retrieved neighbors as semantic priors to guide the generation of latent neighbors, leading to a higher performance.

(2) We observed that MMSBR and DIMO achieved good performance, indicating that the multi-modal information introduced played a positive role in improving the session modeling effect. Diffusion-based models such as DiffuRec and DiQDiff also showed promising results, demonstrating the advantages of diffusion in reconstructing complex distributions and enabling condition-controlled generation. In comparison, the proposed DiffSBR further improves the recommendation performance. DiffSBR generates latent neighbors by continuously injecting both prior knowledge and multi-modal signals into the diffusion-based generation process, thereby exploring the full interest space beyond the observed data.

(3) DiffSBR consistently outperforms all baselines across multiple datasets, demonstrating its effectiveness. This performance gain can be attributed to the crucial role of latent neighbors, as well as the effectiveness of the Retrieval-augmented and Self-augmented Diffusion modules. By leveraging retrieved real neighbors as prior knowledge to guide the generation process and injecting multi-modal signals, DiffSBR is able to generate high-quality latent neighbors beyond the scope of observed data, which strengthens session representations and thereby improving recommendation accuracy. 

\begin{figure}[tb]
	\begin{center}
\includegraphics[width=1\linewidth]{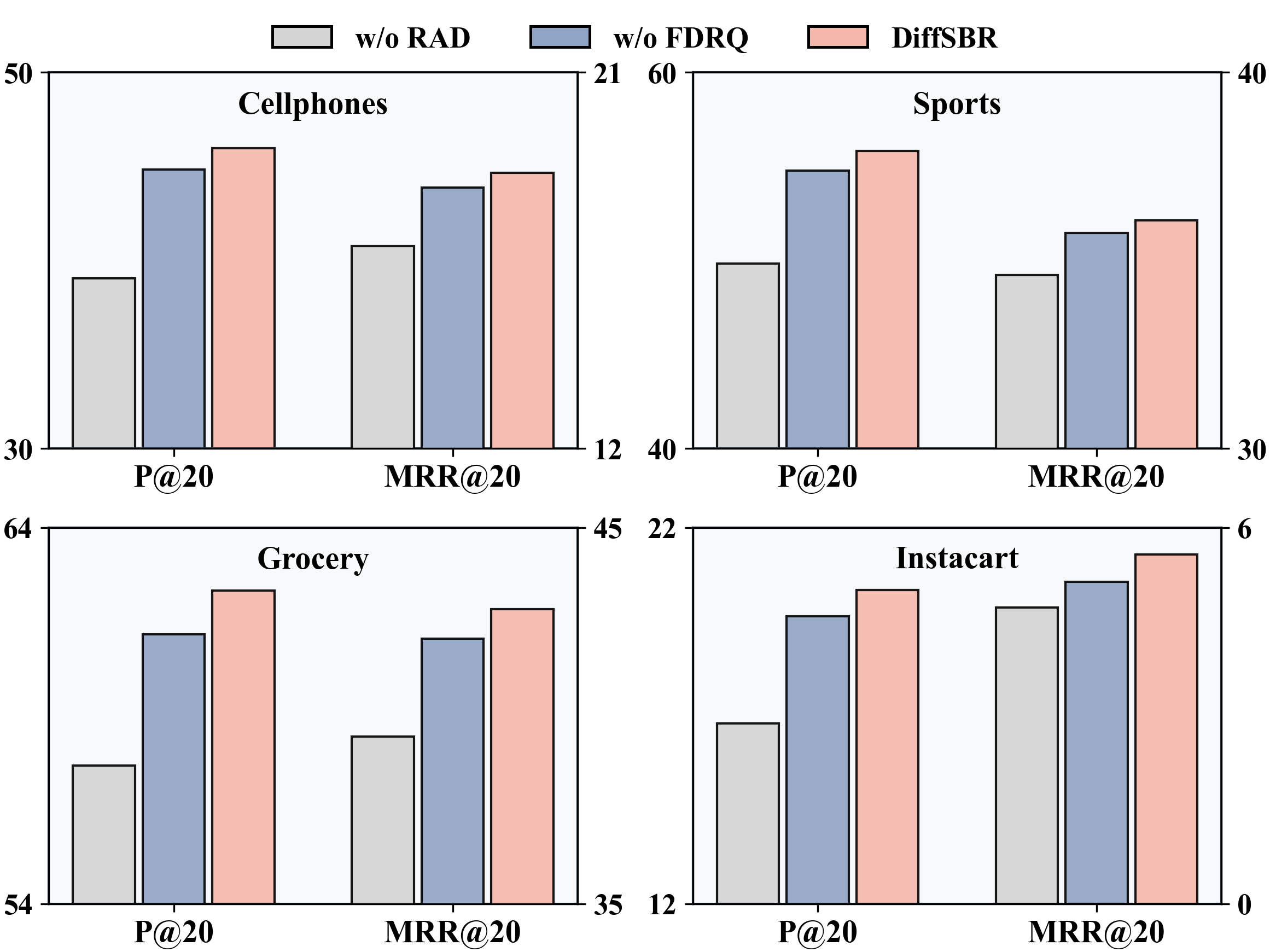}
		
	\end{center}
	\caption{Effect of retrieval-augmented diffusion module.}
	
	\label{fig:ab}
\end{figure}

\subsection{Ablation Studies (RQ2)}
To assess the impact of each major component in our method, we perform ablation studies by selectively removing or altering core modules.
\subsubsection{Effect of Retrieval-augmented Diffusion Module}
To verify the effect of the Retrieval-augmented Diffusion Module,  we compare with
variants: \textbf{w/o RAD}: This variant removes the retrieval-augmented diffusion module, meaning that no retrieved prior information is used to guide the diffusion process, nor is the retriever optimized via feedback, and no multi-modal information is incorporated. \textbf{w/o FDRQ}: In this variant, within the retrieval-augmented diffusion module, only the feedback-driven retriever optimization is removed. The performance variations across four benchmark datasets are illustrated in Figure~\ref{fig:ab}. 

The w/o RAD variant leads to a significant performance drop, indicating the importance of prior knowledge in guiding the diffusion process to generate semantically aligned neighbor representations. Without informative guidance, the diffusion process is more likely to deviate in latent space, resulting in degraded recommendation quality. Moreover, the w/o FDRQ variant weakens the collaboration between the retriever and the generator. Without dynamic supervision from the generator, the retriever cannot adaptively refine its scoring function to better support the generation process, thereby limiting the quality of retrieved guidance signals and lowering the performance.

\subsubsection{Effect of Self-augmented Diffusion Module} We
design these variants: 
\textbf{w/o SAD}: This variant removes the self-augmented diffusion module, meaning that multi-modal information is not utilized in the whole framework and only the ID modality is used. \textbf{-EF}: This variant removes the self-augmentation diffusion module and adopts an early-fusion strategy, where the modality embeddings are directly integrated into the input of the retrieval-augmented diffusion module. \textbf{-LF}: This variant removes the self-augmentation diffusion module and performs late fusion by incorporating modality embeddings at the prediction stage. \textbf{-CF}: This variant removes the self-augmentation diffusion module and performs conditional fusion by directly merging the modality embeddings into the retrieval-augmented diffusion module’s condition, which is then used to guide the diffusion denoising process. \textbf{-AF}: This variant keeps the self-augmentation diffusion module but replaces the contrastive learning with a cross-modal attention mechanism to integrate modality information. 

As shown in Table~\ref{tab:ablation}, for the variant w/o SAD, the performance drops significantly when this variant is removed, indicating that adding multi-modal information to the diffusion process is helpful in improving the quality of latent neighbors. We further evaluate four simplified alternatives (EF, LF, CF, and AF), each integrating multi-modal information through a different fusion strategy. Although these variants incorporate multi-modal information to some extent, they all exhibit clear performance degradation compared to the full model, largely because their fusion mechanisms are either too coarse or insufficiently aligned with the diffusion denoising process. As a result, they fail to provide effective multi-modal signals, leading to weaker latent neighbor quality and reduced model performance. This result highlights that simple fusion strategies are insufficient for effectively leveraging multi-modal signals.

\begin{table}[t]
\centering
% \caption{Ablation study on key components of DiffSBR.}
\caption{Effect of self-augmented diffusion module.}

\label{tab:ablation}
\setlength{\tabcolsep}{4pt}
\renewcommand{\arraystretch}{1.2} 

\resizebox{\columnwidth}{!}{%
  \begin{tabular}{c@{\hspace{2pt}}cc@{\hspace{4pt}}cc@{\hspace{4pt}}cc@{\hspace{4pt}}cc}
    \toprule
    \multirow{2}{*}{\textbf{Method}} & \multicolumn{2}{c}{\textbf{Cellphones}} & \multicolumn{2}{c}{\textbf{Sports}} & \multicolumn{2}{c}{\textbf{Grocery}} & \multicolumn{2}{c}{\textbf{Instacart}} \\
    \cmidrule(lr){2-3} \cmidrule(lr){4-5} \cmidrule(lr){6-7} \cmidrule(lr){8-9}
    & P@20 & MRR@20 & P@20 & MRR@20 & P@20 & MRR@20 & P@20 & MRR@20 \\
    \midrule
    w/o SAD  & 43.19 & 17.76 & 53.90 & 35.43 & 60.70 & 41.73 & 18.55  & 4.97 \\
     -EF & 40.20 & 17.15 & 50.41 & 34.58 & 58.83 & 39.18 & 15.85 & 4.54 \\ 
     -LF & 43.56 & 17.95 & 52.69 & 34.77 & 61.07 & 41.51 & 19.15 & 4.77 \\ 
     -CF & 39.98 & 17.32 & 51.09 & 35.08 & 59.08 & 40.53 & 17.56 & 4.79\\
     -AF & 44.19 & 17.68 & 53.82 & 34.83 & 61.40 & 40.84 & 19.89 & 5.30\\
    \midrule
    \textbf{DiffSBR}      & \textbf{45.97*} & \textbf{18.60*} & \textbf{55.83*} & \textbf{36.07*} & \textbf{62.33*} & \textbf{42.84*} & \textbf{20.34*} & \textbf{5.57*} \\
    \bottomrule
  \end{tabular}  
}
\end{table}

\subsection{Effectiveness of the Generated Latent Neighbors (RQ3)}
To comprehensively assess the necessity and effectiveness of latent neighbor generation, we conduct both quantitative and qualitative analyses. Specifically, we compare two variants: Observable-Neighbors, which retrieves neighbors directly from observed data based on similarity (reflecting traditional retrieval-based methods). Latent-Neighbors, which generates latent neighbors through our proposed DiffSBR approach.

\begin{figure}[tb]
	\begin{center}
\includegraphics[width=1\linewidth]{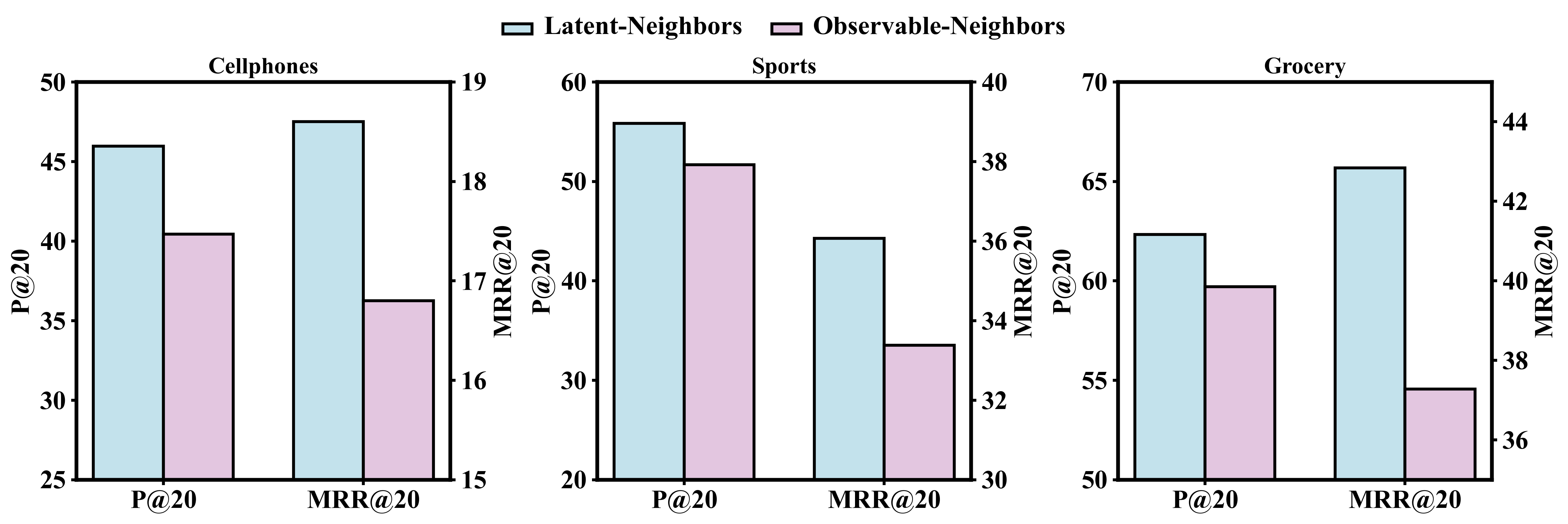}
		
	\end{center}
	\caption{Comparison results of the Latent-Neighbors vs. Observable-Neighbors.}
	
	\label{fig:latent_vs_real}
\end{figure}

\textbf{Quantitative Evaluation.}
To assess the effectiveness of latent neighbor generation, we conduct systematic comparisons between two variants: Latent-Neighbors and Observable-Neighbors across three benchmark datasets. As shown in Figure~\ref{fig:latent_vs_real}\footnote{\label{fn:myfootnote} It is worth noting that we have conducted experiments on four datasets. The observed patterns are consistent with those shown in the Figure. For the sake of visual clarity and presentation aesthetics, we display only a subset of the results, which are representative rather than accidental.}, Latent-Neighbors consistently outperforms Observable-Neighbors in both P@20 and MRR@20 across three datasets. These improvements highlight the advantage of our diffusion-based generation approach, which enables the synthesis of semantically relevant neighbors beyond the observed interaction data. By bridging the gaps left by static retrieval methods, latent neighbors not only enhance overall recommendation performance, but also offer data-driven evidence for the feasibility of generative neighbors modeling.

\begin{figure}[tb]
	\begin{center}
\includegraphics[width=1\linewidth]{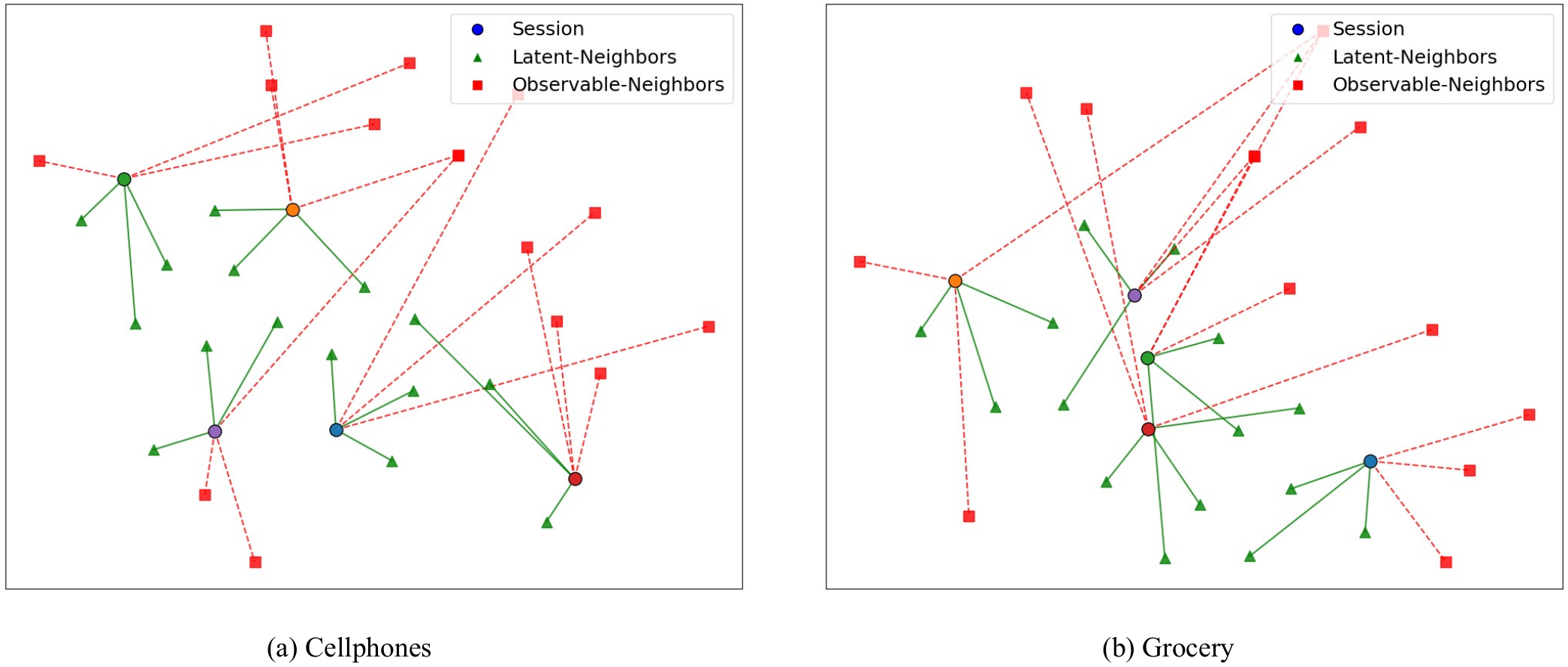}
		
	\end{center}
	\caption{Qualitative visualization of the Latent-Neighbors and Observable-Neighbors.}
	
	\label{fig:tsne}
\end{figure}

\textbf{Qualitative Visualization.}
To intuitively understand why latent neighbors outperform static retrieval ones, we perform a visualization analysis to compare their spatial distribution. Specifically, we project the high-dimensional representations of the target sessions and their corresponding neighbors into a 2D space using t-SNE\cite{mb42,mb53}. We use blue dots to represent different sessions. The red squares in the figure represent the Observable-Neighbors, which are the three neighbors with the highest similarity retrieved from the known dataset. The green triangles in the figure represent the Latent-Neighbors generated using the proposed method.

 As illustrated in Figure~\ref{fig:tsne}\hyperref[fn:myfootnote]{\textsuperscript{\ref*{fn:myfootnote}}}, overall, latent neighbors (denoted by green triangles) exhibit a more concentrated distribution in the embedding space, closely adjacent to the target session node. In contrast, observable neighbors (depicted by red squares) display a comparatively scattered distribution, with some positioned farther from the target session. This phenomenon suggests that while conventional retrieval methods can identify relatively similar neighbors from observed data, the retrieved neighbors are often confined to the scope of recorded behaviors. Conversely, latent neighbors effectively fill the sparse regions of the embedding space, thereby uncovering neighbors that are semantically closer to the target session. These qualitative results substantiate the necessity and effectiveness of our latent neighbor generation approach.

\subsection{Sensitivity Analysis (RQ4)}\label{5.5}

\subsubsection{Impact of the Number of Retrieved Neighbors $k$}

\begin{figure}[tb]
	\begin{center}
		\includegraphics[width=1\linewidth]{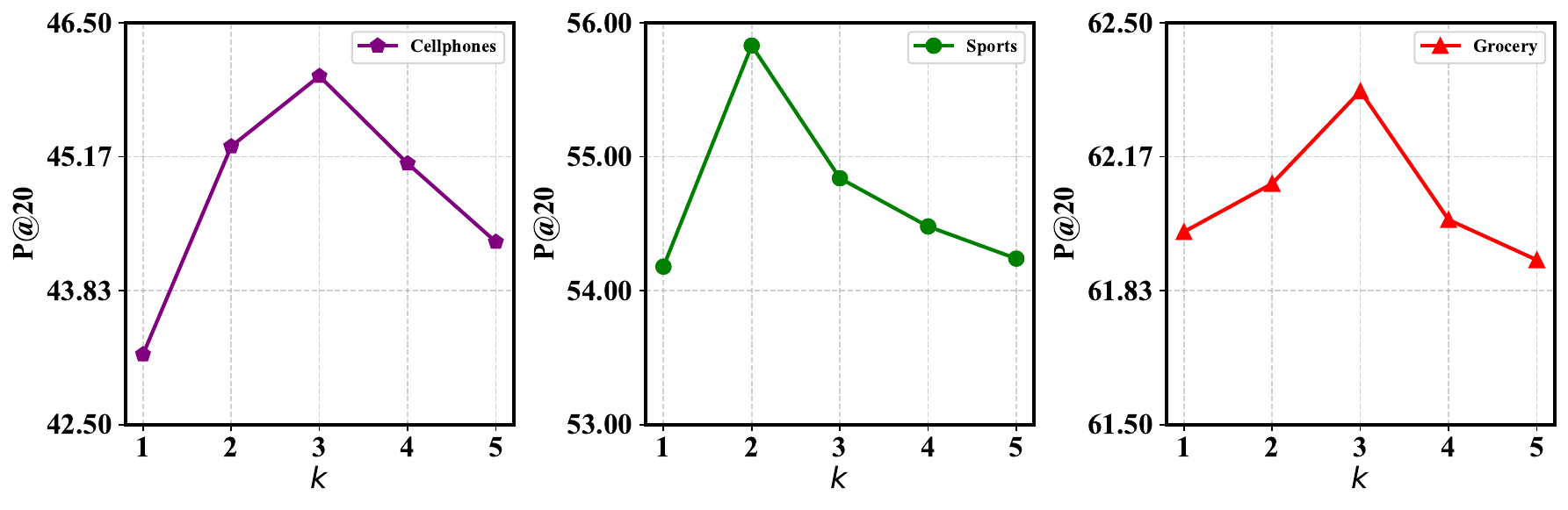}
		
	\end{center}
	\caption{Impact of the number of retrieved neighbors $k$.}
	
	\label{fig:k}
\end{figure}

The number of retrieved neighbors $k$, acting as prior knowledge to guide the generation of latent neighbors, serves as a key hyperparameter.
To investigate its effect, we vary $k$ in $\left \{ 1, 2, 3, 4, 5 \right \} $ and evaluate the model performance across datasets. As shown in Figure~\ref{fig:k}, increasing $k$ initially improves performance, as more informative neighbors provide richer semantic prior signals to the generator. However, performance peaks at $k=2$ for the Sports dataset, at $k=3$ for both Cellphones and Grocery, beyond which further increasing $k$ leads to slight declines. This trend suggests that introducing too many neighbors may inject noise or redundant information, ultimately weakening the effectiveness of guidance.  

\subsubsection{Impact of Loss Weights $\gamma$ and $\delta$}
We further investigate the impact of the two loss weights: the diffusion model loss weight $\gamma$ and contrastive loss weight $\delta$. 
\begin{figure}[tb]
	\begin{center}
\includegraphics[width=0.9\linewidth]{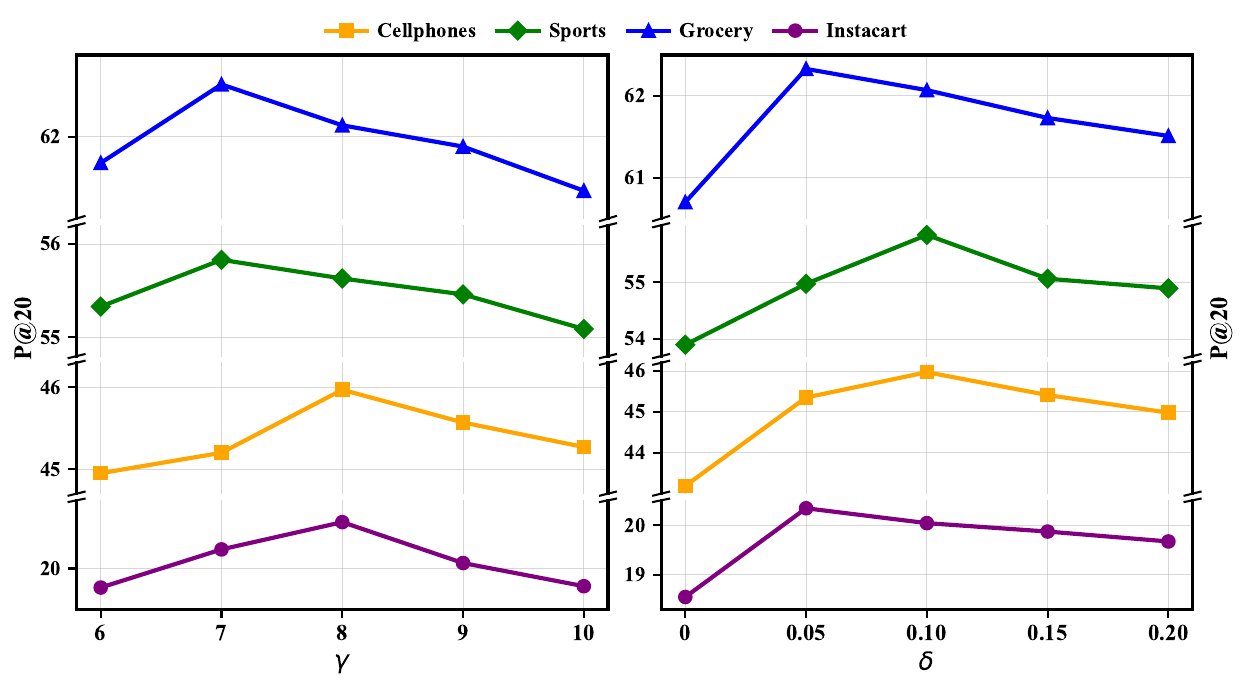}
		
	\end{center}
	\caption{Impact of loss weights $\gamma$ and $\delta$.}
	 
	\label{fig:diff}
\end{figure}
As shown in Figure~\ref{fig:diff}, for the diffusion loss weight $\gamma$, we analyze the results from $\gamma=6$, as preliminary experiments indicated that smaller values (e.g., $\gamma<6$) led to suboptimal performance due to insufficient influence of the diffusion loss on the generation process. Specifically, the Sports and Grocery datasets reach their peak performance at $\gamma=7$, while the Cellphones and Instacart datasets perform best at $\gamma=8$. This suggests that an appropriate weighting of the diffusion model loss allows the model to effectively generate latent neighbors. However, setting $\gamma$ too high (e.g., $\gamma=9$ or 10) results in a slight decline in performance, likely because the diffusion loss term dominates the overall objective and diminishes the contribution of other important components.  

For the contrastive loss weight $\delta$, performance improves with a moderate contrastive signal. Specifically, the Grocery and Instacart datasets achieve their best performance at $\delta = 0.05$, while the Cellphones and Sports datasets reach their peak at $\delta = 0.1$. This confirms the effectiveness of injecting multi-modal information during the diffusion process. However, excessively large $\delta$ values may interfere with the optimization of other loss components, leading to slightly reduced performance. 

\section{Conclusion}
In this study, we propose a novel Diffusion-based Latent Neighbor Generation model for improving SBR. Specifically, we design a retrieval-augmented diffusion module that leverages retrieved neighbors as prior knowledge to guide the diffusion process in generating latent neighbors. Within this module, we further introduce a new training strategy to enhance the synergy between retrieval and diffusion during neighbor generation. In addition, we develop a self-augmented diffusion module to fully exploit multi-modal information throughout the diffusion process, thereby improving the quality of generated neighbors. Experimental results on four benchmark datasets demonstrate that DiffSBR consistently achieves significant performance gains over state-of-the-art methods.

In this work, we highlight the potential of latent neighbor generation for SBR. However, this remains a preliminary study on modeling latent neighbors, and there is still room for improvement. 
In future work, we plan to investigate the distinct strengths and complementarities of various generative models (e.g., autoregressive models) and explore hybrid frameworks that integrate multiple generative paradigms, aiming to further enhance the quality of generated latent neighbors.

\begin{acks}
This research was supported by the National Natural Science Foundation of China (62402093), and the Sichuan Science and Technology Program (2025ZNSFSC0479).
\end{acks}

\bibliographystyle{ACM-Reference-Format}
\balance
\bibliography{sample-base}

@String{Computing = "Computing" }

@inproceedings{mb01,
  title={DiffMM: Multi-modal diffusion model for recommendation},
  author={Jiang, Yangqin and Xia, Lianghao and Wei, Wei and Luo, Da and Lin, Kangyi and Huang, Chao},
  booktitle={MM},
  pages={7591--7599},
  year={2024}
}

@inproceedings{mb02,
  title={Distinguished quantized guidance for diffusion-based sequence recommendation},
  author={Mao, Wenyu and Liu, Shuchang and Liu, Haoyang and Liu, Haozhe and Li, Xiang and Hu, Lantao},
  booktitle={WWW},
  pages={425--435},
  year={2025}
}

@article{mb03,
  title={Enhancing collaborative information with contrastive learning for session-based recommendation},
  author={An, Guojia and Sun, Jing and Yang, Yuhan and Sun, Fuming},
  journal={Information Processing \& Management (IPM)},
  volume={61},
  pages={103738},
  year={2024},
}

@article{mb04,
  title={Beyond co-occurrence: Multi-modal session-based recommendation},
  author={Zhang, Xiaokun and Xu, Bo and Ma, Fenglong and Li, Chenliang and Yang, Liang and Lin, Hongfei},
  journal={Transactions on Knowledge and Data Engineering (TKDE)},
  volume={36},
  pages={1450--1462},
  year={2023},

}

@inproceedings{mb05,
  title={Dynamic multi-interest graph neural network for session-based recommendation},
  author={Lv, Mingyang and Liu, Xiangfeng and Xu, Yuanbo},
  booktitle={AAAI},
  year={2025},
  pages={12328--12336},
  
}

@inproceedings{mb06,
  title={Bi-channel multiple sparse graph attention networks for session-based recommendation},
  author={Qiao, Shutong and Zhou, Wei and Wen, Junhao and Zhang, Hongyu and Gao, Min},
  booktitle={CIKM},
  pages={2075--2084},
  year={2023}
}

@article{mb07,
  title={Dynamic intent-aware iterative denoising network for session-based recommendation},
  author={Zhang, Xiaokun and Lin, Hongfei and Xu, Bo and Li, Chenliang and Lin, Yuan and Liu, Haifeng and Ma, Fenglong},
  journal={Information Processing \& Management (IPM)},
  volume={59},
  pages={102936},
  year={2022},
}

@inproceedings{mb08,
  title={An intent-guided collaborative machine for session-based recommendation},
  author={Pan, Zhiqiang and Cai, Fei and Ling, Yanxiang and De Rijke, Maarten},
  booktitle={SIGIR},
  pages={1833--1836},
  year={2020}
}

@article{mb09,
  title={Collaborative graph learning for session-based recommendation},
  author={Zhiqiang, Pan and Fei, Cai and Wanyu, Chen and Chonghao, Chen and Honghui, Chen},
  journal={Transactions on Information System (TOIS)},
  volume={40},
  pages={72},
  year={2022}
}

@inproceedings{mb10,
  title={Disentangling id and modality effects for session-based recommendation},
  author={Zhang, Xiaokun and Xu, Bo and Ren, Zhaochun and Wang, Xiaochen and Lin, Hongfei and Ma, Fenglong},
  booktitle={SIGIR},
  pages={1883--1892},
  year={2024}
}

@article{mb11,
  title={Noise-reducing graph neural network with intent-target co-action for session-based recommendation},
  author={Qiao, Shutong and Zhou, Wei and Luo, Fengji and Wen, Junhao},
  journal={Information Processing \& Management (IPM)},
  volume={60},
  pages={103517},
  year={2023},
}

@inproceedings{mb12,
  title={Global context enhanced graph neural networks for session-based recommendation},
  author={Wang, Ziyang and Wei, Wei and Cong, Gao and Li, Xiao-Li and Mao, Xian-Ling and Qiu, Minghui},
  booktitle={SIGIR},
  pages={169--178},
  year={2020}
}

@inproceedings{mb13,
  title={Price does matter! modeling price and interest preferences in session-based recommendation},
  author={Zhang, Xiaokun and Xu, Bo and Yang, Liang and Li, Chenliang and Ma, Fenglong and Liu, Haifeng and Lin, Hongfei},
  booktitle={SIGIR},
  pages={1684--1693},
  year={2022}
}

@inproceedings{mb14,
  title={Neural attentive session-based recommendation},
  author={Li, Jing and Ren, Pengjie and Chen, Zhumin and Ren, Zhaochun and Lian, Tao and Ma, Jun},
  booktitle={CIKM},
  pages={1419--1428},
  year={2017}
}

@inproceedings{mb15,
  title={Session-based recommendation with graph neural networks},
  author={Wu, Shu and Tang, Yuyuan and Zhu, Yanqiao and Wang, Liang and Xie, Xing and Tan, Tieniu},
  booktitle={AAAI},
  pages={346--353},
  year={2019}
}

@inproceedings{mb16,
  title={An attribute-driven mirror graph network for session-based recommendation},
  author={Lai, Siqi and Meng, Erli and Zhang, Fan and Li, Chenliang and Wang, Bin and Sun, Aixin},
  booktitle={SIGIR},
  pages={1674--1683},
  year={2022}
}

@inproceedings{mb17,
  title={Learning multi-granularity consecutive user intent unit for session-based recommendation},
  author={Guo, Jiayan and Yang, Yaming and Song, Xiangchen and Zhang, Yuan and Wang, Yujing and Bai, Jing and Zhang, Yan},
  booktitle={WSDM},
  pages={343--352},
  year={2022}
}

@inproceedings{mb18,
  title={Efficiently leveraging multi-level user intent for session-based recommendation via atten-mixer network},
  author={Zhang, Peiyan and Guo, Jiayan and Li, Chaozhuo and Xie, Yueqi and Kim, Jae Boum and Zhang, Yan and Xie, Xing and Wang, Haohan and Kim, Sunghun},
  booktitle={WSDM},
  pages={168--176},
  year={2023}
}

@inproceedings{mb19,
  title={A dynamic recurrent model for next basket recommendation},
  author={Yu, Feng and Liu, Qiang and Wu, Shu and Wang, Liang and Tan, Tieniu},
  booktitle={SIGIR},
  pages={729--732},
  year={2016}
}

@article{mb20,
  title={Session-based recommendations with recurrent neural networks},
  author={Hidasi, Bal{\'a}zs and Karatzoglou, Alexandros and Baltrunas, Linas and Tikk, Domonkos},
  journal={arXiv preprint arXiv:1511.06939},
  year={2015}
}

@inproceedings{mb21,
  title={TAGNN: Target attentive graph neural networks for session-based recommendation},
  author={Yu, Feng and Zhu, Yanqiao and Liu, Qiang and Wu, Shu and Wang, Liang and Tan, Tieniu},
  booktitle={SIGIR},
  pages={1921--1924},
  year={2020}
}

@article{mb22,
  title={Exploiting cross-session information for session-based recommendation with graph neural networks},
  author={Qiu, Ruihong and Huang, Zi and Li, Jingjing and Yin, Hongzhi},
  journal={Transactions on Information Systems (TOIS)},
  volume={38},
  pages={1--23},
  year={2020},
}

@inproceedings{mb23,
  title={Exploiting explicit and implicit item relationships for session-based recommendation},
  author={Li, Zihao and Wang, Xianzhi and Yang, Chao and Yao, Lina and McAuley, Julian and Xu, Guandong},
  booktitle={WSDM},
  pages={553--561},
  year={2023}
}

@article{mb24,
  title={Multi-view Intent Learning and Alignment with Large Language Models for Session-based Recommendation},
  author={Qiao, Shutong and Zhou, Wei and Wen, Junhao and Gao, Chen and Luo, Qun and Chen, Peixuan and Li, Yong},
  journal={Transactions on Information Systems (TOIS)},
  volume={43},
  pages={1--25},
  year={2025},
 
}

@inproceedings{mb25,
  title={Diffusion recommender model},
  author={Wang, Wenjie and Xu, Yiyan and Feng, Fuli and Lin, Xinyu and He, Xiangnan and Chua, Tat-Seng},
  booktitle={SIGIR},
  pages={832--841},
  year={2023}
}

@article{mb26,
  title={Diffurec: A diffusion model for sequential recommendation},
  author={Li, Zihao and Sun, Aixin and Li, Chenliang},
  journal={Transactions on Information Systems (TOIS)},
  volume={42},
  pages={1--28},
  year={2023},
}

@inproceedings{mb27,
  title={Diff4rec: Sequential recommendation with curriculum-scheduled diffusion augmentation},
  author={Wu, Zihao and Wang, Xin and Chen, Hong and Li, Kaidong and Han, Yi and Sun, Lifeng and Zhu, Wenwu},
  booktitle={MM},
  pages={9329--9335},
  year={2023}
}

@inproceedings{mb28,
  title={Diffusion augmentation for sequential recommendation},
  author={Liu, Qidong and Yan, Fan and Zhao, Xiangyu and Du, Zhaocheng and Guo, Huifeng and Tang, Ruiming and Tian, Feng},
  booktitle={CIKM},
  pages={1576--1586},
  year={2023}
}

@inproceedings{mb29,
  title={Diffkg: Knowledge graph diffusion model for recommendation},
  author={Jiang, Yangqin and Yang, Yuhao and Xia, Lianghao and Huang, Chao},
  booktitle={WSDM},
  pages={313--321},
  year={2024}
}

@inproceedings{mb30,
  title={Denoising diffusion recommender model},
  author={Zhao, Jujia and Wenjie, Wang and Xu, Yiyan and Sun, Teng and Feng, Fuli and Chua, Tat-Seng},
  booktitle={SIGIR},
  pages={1370--1379},
  year={2024}
}

@inproceedings{mb31,
  title={Multimodal conditioned diffusion model for recommendation},
  author={Ma, Haokai and Yang, Yimeng and Meng, Lei and Xie, Ruobing and Meng, Xiangxu},
  booktitle={WWW},
  pages={1733--1740},
  year={2024}
}

@inproceedings{mb32,
  title={Denoising diffusion probabilistic models},
  author={Ho, Jonathan and Jain, Ajay and Abbeel, Pieter},
  booktitle={NeurIPS},
  pages={6840--6851},
  year={2020}
}

@article{mb33,
  title={Diffusion models: A comprehensive survey of methods and applications},
  author={Yang, Ling and Zhang, Zhilong and Song, Yang and Hong, Shenda and Xu, Runsheng and Zhao, Yue and Zhang, Wentao and Cui, Bin and Yang, Ming-Hsuan},
  journal={Computing Surveys (CSUR)},
  volume={56},
  pages={1--39},
  year={2023},
}

@inproceedings{mb34,
  title={Patch diffusion: Faster and more data-efficient training of diffusion models},
  author={Wang, Zhendong and Jiang, Yifan and Zheng, Huangjie and Wang, Peihao and He, Pengcheng and Wang, Zhangyang and Chen, Weizhu and Zhou, Mingyuan and others},
  booktitle={NeurIPS},
  pages={72137--72154},
  year={2023}
}

@inproceedings{mb35,
  title={Going deeper with convolutions},
  author={Szegedy, Christian and Liu, Wei and Jia, Yangqing and Sermanet, Pierre and Reed, Scott and Anguelov, Dragomir and Erhan, Dumitru and Vanhoucke, Vincent and Rabinovich, Andrew},
  booktitle={CVPR},
  pages={1--9},
  year={2015}
}

@inproceedings{mb36,
  title={Bert: Pre-training of deep bidirectional transformers for language understanding},
  author={Devlin, Jacob and Chang, Ming-Wei and Lee, Kenton and Toutanova, Kristina},
  booktitle={NAACL-HLT},
  pages={4171--4186},
  year={2019}
}

@inproceedings{mb37,
  title={Self-supervised graph co-training for session-based recommendation},
  author={Xia, Xin and Yin, Hongzhi and Yu, Junliang and Shao, Yingxia and Cui, Lizhen},
  booktitle={CIKM},
  pages={2180--2190},
  year={2021}
}

@inproceedings{mb38,
  title={Generate what you prefer: Reshaping sequential recommendation via guided diffusion},
  author={Yang, Zhengyi and Wu, Jiancan and Wang, Zhicai and Wang, Xiang and Yuan, Yancheng and He, Xiangnan},
  booktitle={NeurIPS},
  pages={24247--24261},
  year={2023}
}

@inproceedings{mb39,
  title={Improving Retrieval-Augmented Code Comment Generation by Retrieving for Generation},
  author={Lu, Hanzhen and Liu, Zhongxin},
  booktitle={ICSME},
  pages={350--362},
  year={2024},
}

@inproceedings{mb40,
  title={When recurrent neural networks meet the neighborhood for session-based recommendation},
  author={Jannach, Dietmar and Ludewig, Malte},
  booktitle={RecSys},
  pages={306--310},
  year={2017}
}

@inproceedings{mb41,
  title={Coral: collaborative retrieval-augmented large language models improve long-tail recommendation},
  author={Wu, Junda and Chang, Cheng-Chun and Yu, Tong and He, Zhankui and Wang, Jianing and Hou, Yupeng and McAuley, Julian},
  booktitle={SIGKDD},
  pages={3391--3401},
  year={2024}
}

@article{mb42,
  title={Visualizing data using t-SNE},
  author={Van der Maaten, Laurens and Hinton, Geoffrey},
  journal={Journal of Machine Learning Research (JMLR)},
  volume={9},
  number={11},
  year={2008}
}

@inproceedings{mb43,
  title={Beyond whole dialogue modeling: Contextual disentanglement for conversational recommendation},
  author={An, Guojia and Zou, Jie and Wei, Jiwei and Zhang, Chaoning and Sun, Fuming and Yang, Yang},
  booktitle={SIGIR},
  pages={31--41},
  year={2025}
}

@inproceedings{mb45,
  title={Hierarchical Intent-guided Optimization with Pluggable LLM-Driven Semantics for Session-based Recommendation},
  author={Chen, Jinpeng and He, Jianxiang and Li, Huan and Wang, Senzhang and Cao, Yuan and Wei, Kaimin and Yang, Zhenye and Ji, Ye},
  booktitle={SIGIR},
  pages={1655--1665},
  year={2025}
}

@inproceedings{mb46,
  title={Self-supervised hypergraph convolutional networks for session-based recommendation},
  author={Xia, Xin and Yin, Hongzhi and Yu, Junliang and Wang, Qinyong and Cui, Lizhen and Zhang, Xiangliang},
  booktitle={AAAI},
  pages={4503--4511},
  year={2021}
}

@inproceedings{mb47,
  title={Generative adversarial nets},
  author={Goodfellow, Ian J and Pouget-Abadie, Jean and Mirza, Mehdi and Xu, Bing and Warde-Farley, David and Ozair, Sherjil and Courville, Aaron and Bengio, Yoshua},
  booktitle={NeurIPS},
  pages={1--9},
  year={2014}
}

@inproceedings{mb48,
  title={Variational autoencoder for deep learning of images, labels and captions},
  author={Pu, Yunchen and Gan, Zhe and Henao, Ricardo and Yuan, Xin and Li, Chunyuan and Stevens, Andrew and Carin, Lawrence},
  booktitle={NeurIPS},
  pages={1--9},
  year={2016}
}

@inproceedings{mb49,
  title={Diffusion models beat gans on image synthesis},
  author={Dhariwal, Prafulla and Nichol, Alexander},
  booktitle={NeurIPS},
  pages={8780--8794},
  year={2021}
}

@article{mb50,
  title={Guided Diffusion-based Counterfactual Augmentation for Robust Session-based Recommendation},
  author={Gupta, Muskan and Gupta, Priyanka and Vig, Lovekesh},
  journal={arXiv preprint arXiv:2410.21892},
  year={2024}
}

@inproceedings{mb51,
  title={Towards question-based recommender systems},
  author={Zou, Jie and Chen, Yifan and Kanoulas, Evangelos},
  booktitle={SIGIR},
  pages={881--890},
  year={2020}
}

@inproceedings{mb52,
  title={PSCon: Product Search Through Conversations},
  author={Zou, Jie and Aliannejadi, Mohammad and Kanoulas, Evangelos and Han, Shuxi and Ma, Heli and Wang, Zheng and Yang, Yang and Shen, Heng Tao},
  booktitle={SIGIR},
  pages={3659--3669},
  year={2025}
}

@inproceedings{mb53,
  title={Adversarial cross-modal retrieval},
  author={Wang, Bokun and Yang, Yang and Xu, Xing and Hanjalic, Alan and Shen, Heng Tao},
  booktitle={MM},
  pages={154--162},
  year={2017}
}

@article{mb54,
  title={Geometric matching for cross-modal retrieval},
  author={Wang, Zheng and Gao, Zhenwei and Yang, Yang and Wang, Guoqing and Jiao, Chengbo and Shen, Heng Tao},
  journal={Transactions on Neural Networks and Learning Systems (TNNLS)},
  volume={36},
  number={3},
  pages={5509--5521},
  year={2024},
}

@inproceedings{mb55,
  title={Retrieval Augmented Multi-agent Recommender},
  author={Ao, Xiao and Han, Shuxi and Li, Yeming and Ma, Heli and Zhang, Pengfei and Zou, Jie},
  booktitle={WWW},
  pages={2968--2972},
  year={2025}
}

@article{mb56,
  title={Video captioning with attention-based LSTM and semantic consistency},
  author={Gao, Lianli and Guo, Zhao and Zhang, Hanwang and Xu, Xing and Shen, Heng Tao},
  journal={Transactions on Multimedia (TMM)},
  volume={19},
  number={9},
  pages={2045--2055},
  year={2017},
}

@inproceedings{mb57,
  title={MSCRS: Multi-modal semantic graph prompt learning framework for conversational recommender systems},
  author={Wei, Yibiao and Zou, Jie and Guo, Weikang and Wang, Guoqing and Xu, Xing and Yang, Yang},
  booktitle={SIGIR},
  pages={42--52},
  year={2025}
}

\end{document}